\documentclass[preprint,journal]{vgtc}       

\ifpdf
  \pdfoutput=1\relax                   
  \usepackage{graphicx}                
  \DeclareGraphicsExtensions{.pdf,.png,.jpg,.jpeg} 
\else
  \usepackage{graphicx}                
  \DeclareGraphicsExtensions{.eps}     
\fi%

\graphicspath{{figures/}{pictures/}{images/}{./}} 

\usepackage{microtype}                 
\PassOptionsToPackage{warn}{textcomp}  
\usepackage{textcomp}                  
\usepackage{mathptmx}                  
\usepackage{times}                     
\usepackage{cite}                      
\usepackage{tabu}                      
\usepackage{booktabs}                  
\usepackage[shortlabels]{enumitem}
\usepackage{multirow}
\usepackage{float}
\usepackage{array}

\preprinttext{This is the author’s version of the article.}

\onlineid{0}

\vgtccategory{Research}
\vgtcpapertype{please specify}

\title{Working in Extended Reality in the Wild: Worker and Bystander Experiences of XR Virtual Displays in Real-World Settings}

\author{Leonardo Pavanatto$^*$, Verena Biener$^*$, Jennifer Chandran, Snehanjali Kalamkar, Feiyu Lu, John J. Dudley,\\Jinghui Hu, G. Nikki Ramirez-Saffy, Per Ola Kristensson, Alexander Giovannelli, Luke Schlueter, Jörg Müller,\\Jens Grubert, and Doug A. Bowman} 
\authorfooter{
\item
 Leonardo Pavanatto, Jennifer Chandran, Feiyu Lu, Gabriella Nicole Ramirez, Alexander Giovannelli, Luke Schlueter, and Doug A. Bowman are with the Center for Human-Computer Interaction, Department of Computer Science, Virginia Tech, Blacksburg, VA, United States. E-mail: \{lpavanat $|$ jenniferchandran $|$ feiyulu $|$ gnramirezsaffy $|$ agiovannelli $|$ lukeschlueter $|$ dbowman\}@vt.edu
\item 
 Verena Biener is with the Visualization Research Center, University of Stuttgart, Germany. Email: verena.biener@visus.uni-stuttgart.de
 \item 
 Snehanjali Kalamkar and Jens Grubert are with Coburg University of Applied Sciences and Arts, Germany. Email: \{snehanjali.kalamkar $|$ jens.grubert\}@hs-coburg.de
 \item 
 Jörg Müller is with the University of Bayreuth, Germany. Email: joerg.mueller@uni-bayreuth.de
 \item 
 John J. Dudley, Jinghui Hu and Per Ola Kristensson are with the Department of Engineering, University of Cambridge, UK. Email: \{jjd50 $|$ jh2265 $|$ pok21\}@cam.ac.uk
\item $^{*}$ Leonardo Pavanatto and Verena Biener contributed equally to this work.
}

\shortauthortitle{? \MakeLowercase{\textit{et al.}}: Performing Knowledge Work from Anywhere}

\abstract{
Although access to sufficient screen space is crucial to knowledge work, workers often find themselves with limited access to display infrastructure in remote or public settings. While virtual displays can be used to extend the available screen space through extended reality (XR) head-worn displays (HWD), we must better understand the implications of working with them in public settings from both users' and bystanders' viewpoints. To this end, we conducted two user studies. We first explored the usage of a hybrid AR display across real-world settings and tasks. We focused on how users take advantage of virtual displays and what social and environmental factors impact their usage of the system. A second study investigated the differences between working with a laptop, an AR system, or a VR system in public. We focused on a single location and participants performed a predefined task to enable direct comparisons between the conditions while also gathering data from bystanders. The combined results suggest a positive acceptance of XR technology in public settings and show that virtual displays can be used to accompany existing devices. We highlighted some environmental and social factors. We saw that previous XR experience and personality can influence how people perceive the use of XR in public. In addition, we confirmed that using XR in public still makes users stand out and that bystanders are curious about the devices, yet have no clear understanding of how they can be used.

} 


\CCScatlist{
  \CCScatTwelve{Human-centered computing}{Human computer interaction (HCI)}{Interaction paradigms}{Mixed / augmented reality};
  \CCScatTwelve{Human-centered computing}{Human computer interaction (HCI)}{Interaction paradigms}{Empirical studies in interaction design}
}

\teaser{
  \centering
  \includegraphics[width=\linewidth]{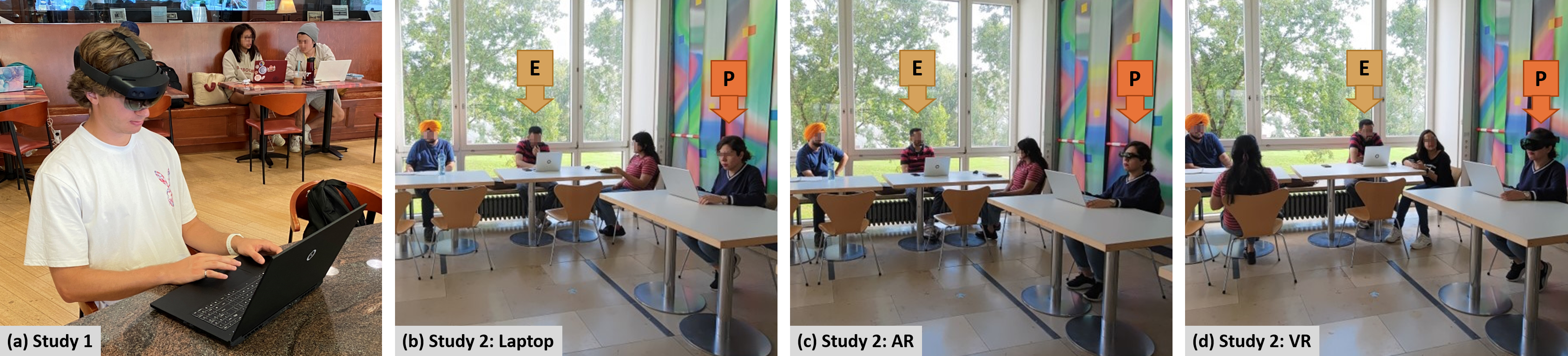}
  \caption{Participant as seen by people around them in study 1 (a), in study 2 during the laptop condition (b), the AR condition (c) and the VR condition (d). E indicates the experimenter of the study and P indicates the participant.}
  \label{fig:teaser}
}

\vgtcinsertpkg

\begin{document}
\firstsection{Introduction}
\maketitle
Knowledge workers often require large display spaces to view, process, and cross-reference content efficiently, which they may not have access to when working remotely. Such workers might be in crowded or confined spaces with many distractions \cite{grubert_office_2018}, relying on the portable devices available to them at the time, such as laptops, tablets, and smartphones \cite{yuan_understanding_2022}. Remote work scenarios arise from necessity (e.g., during travel, or a meeting), inspiration (e.g., being outdoors in proximity to nature), or simply from preference or health considerations (e.g., working from home). Despite being away from their physical offices, these workers still handle complex tasks that demand extensive screen space. Managing many windows on a small screen can be time-consuming \cite{czerwinski_toward_2003}, and it is not unusual for users to forget where certain windows are located since they can be hidden or occluded.

Using extended reality (XR) for knowledge work has gained traction with the promotion of commercial off-the-shelf devices. Prior research has demonstrated the potential benefits of XR for knowledge work, such as how virtual reality (VR) can be used in open office spaces to reduce distractions \cite{ruvimova_transport_2020} and how new interaction possibilities provided by VR can improve the performance for certain tasks \cite{gesslein_pen-based_2020, biener2022povrpoint}. Additionally, researchers have evaluated how the large display space provided by augmented reality (AR) and VR can be used in the context of knowledge work \cite{mcgill_expanding_2020, pavanatto_monitors_2021}. Such virtual displays use a head-worn display (HWD) to render information and applications from a personal computer without being constrained to physical monitors, offering more flexibility (how to display information), portability (where to display information), and scalability (how much information can be displayed) \cite{pavanatto_monitors_2021}.

Existing research has validated the feasibility of performing knowledge work on virtual displays with current technology. Although they do not yet match the performance of physical multi-monitor setups \cite{pavanatto_monitors_2021, biener_quantifying_2022}, virtual displays offer significant advantages over single-screen laptops by eliminating context switches and facilitating head and eye glances \cite{pavanatto_virtual_2023}. However, these studies have predominantly focused on laboratory-based experiments, not considering the unique and ecologically valid public settings where laptops are frequently used. Abowd and Mynatt \cite{abowd_charting_2000} highlighted that everyday computing activities could differ significantly from controlled evaluations in the laboratory, which rarely have clear beginnings/ends, often experience interruptions, and typically involve multiple concurrent activities. Thus, in-the-wild evaluations are crucial to understanding users' subjective 
experiences, including their emotional state and social interactions with people around them. 

In this paper, we conducted two user studies to understand (1) how users take advantage of virtual displays when only a single laptop monitor is available to them, (2) their user experience of working on virtual displays in real-world settings, and (3) public perception of using XR in public spaces. Our first study explored the usage of a hybrid AR see-through system, which extended the screen from a physical laptop to the virtual space. Focusing on ecological validity, we explored the use of this system in four different campus settings, and participants were performing their own self-defined tasks. In contrast, in our second study, participants completed three sessions of working in the same public location, a university cafeteria, with three different systems: a standard laptop, a laptop combined with an AR headset, and a laptop combined with a VR headset. To better compare the systems, this time participants were performing the same kind of task in each session.

\section{Related Work}
\subsection{Pervasive and Everyday XR}
While AR has historically been used to solve domain-specific issues \cite{feiner_touring_1997, thomas_wearable_1998, schall_virtual_2008, henderson_evaluating_2009, collins_computer-assisted_2014}, recent research has explored pervasive and everyday AR \cite{grubert_towards_2017, bellgardt_utilizing_2017, lu_glanceable_2020}. Grubert et al. \cite{grubert_office_2018} introduced pervasive AR, emphasizing general-purpose, all-day use of AR HWDs as a continuous, universal augmented interface. Users could continuously benefit from AR glasses in tasks like information acquisition, productivity, and entertainment. The form factor of current XR devices remains a barrier to adoption as headsets due to stress, mental overload, visual and muscle fatigue, and physical discomfort \cite{souchet2022narrative, guo2020exploring}. Biener et al.~\cite{biener_quantifying_2022} found that using a VR headset resulted in worse ratings than a physical environment for a range of measures such as task load, usability, flow, frustration and visual fatigue. An in-the-wild study by Lu et al.~\cite{lu_evaluating_2021} pointed out that participants could imagine using the AR system daily, if the form factor of the devices would be improved. This work explores how AR displays can enhance everyday productivity and addresses the associated technological and adoption challenges.

\subsection{Social Acceptability of AR/VR in Public Spaces}
An interface with high social acceptability can increase others' interest in including the technology user in their social groups \cite{schwind2018need} or can align with or positively alter the user's self-image and external image \cite{koelle2020social}. Strategies to enhance social acceptability include subtlety, designing devices to look like non-digital accessories, and informing bystanders of the user's activities.

Social acceptability of HWDs in public has been investigated \cite{McGill_2020_Challenges, Bajorunaite_2021}. Williamson et al. \cite{Williamson_2019} noted that VR usage in public spaces might attract undesired attention and recommended maintaining users' peripheral awareness and transitions between AR and VR. Users were found to prefer displays organized vertically to avoid ``social collision’’ on airplanes \cite{ng_passenger_2021} and avoided placing virtual displays over bystanders' faces, bodies, and possessions to maintain situational awareness \cite{medeiros_shielding_2022}. Vergari et al. \cite{vergari2021influence} simulated different social environments with 360° videos to evaluate the effects of using VR in these environments. George et al. \cite{george2019should} explored in a lab study if bystanders can identify task switches of VR users and which strategies they employ to interrupt them.

Field studies in public places include evaluating interactive glasses for notifications while walking through a calm and busy street \cite{lucero2014notifeye}, exploring different interaction techniques and form factors of smart glasses in a cafe\cite{tung2015user}, and elicitation studies for socially acceptable hand-to-face input in busy public spaces \cite{lee2018designing}. Eghbali et al. \cite{eghbali2019social} found VR users' experiences can be influenced by factors such as freedom to switch between realities, sense of safety, physical privacy, and uninterruptible immersion, while bystanders' experience depends on feeling normal, safe, and having a shared experience.

Various questionnaires have been used to measure social acceptability. Kelly et al. \cite{kelly2016wear} proposed the WEAR scale measuring the social acceptability of wearable devices. Ahlström et al. \cite{ahlstrom2014you} focused on users' perceptions, asking them to rate their overall impression during the task and in which situations they would feel comfortable performing the proposed gestures. Koelle et al. \cite{koelle_dont_2015} asked participants to rate their feelings while using the device on pairs of opposites, such as threatened and safe or unsure and self-confident. Similarly, Verma et al. \cite{verma2015study} asked participants to rate their feelings on Likert scales ranging from embarrassed to comfortable and from foolish to sensible.

\subsection{In-The-Wild Evaluation of AR/VR systems}
In-the-wild evaluation refers to experiments with real users in uncontrolled settings \cite{brown_into_2011}. Rather than completing ``arbitrary tasks decided by the researcher'' \cite{rogers_interaction_2011}, participants can understand and appropriate the technology to their needs \cite{rogers_interaction_2011}. Those results may differ from laboratory settings, considering the unpredictability of people, as participants can become distracted, be interrupted, or interact with other people \cite{rogers_interaction_2011}. In-the-wild AR/VR studies have shown their viability and data quality in diverse areas such as education \cite{petersen_pedagogical_2021}, accessibility \cite{schmelter_towards_2023}, recreational sports \cite{colley_skiing_2015, colley_hedonic_2017}, locomotion \cite{ragozin_mazerunvr_2020}, military training \cite{laviola_using_2015}, and perception studies \cite{arora_thinking_2021}. However, there has been limited exploration of how AR HWDs could better facilitate everyday workflows. To our knowledge, our work is one of the first to address this gap, by exploring the uses of AR HWDs to facilitate knowledge work in real-world settings.

\subsection{Extended Reality for Knowledge Work}
Advances in XR technologies have the potential for adoption and benefiting knowledge workers \cite{bellgardt_utilizing_2017, mcguffin_augmented_2019}. We can combine physical and virtual content as if they were the same \cite{li_holodoc_2019}, use immersive systems to facilitate multi-tasking \cite{ens_personal_2014} and sensemaking \cite{kobayashi_translating_2021, lisle_evaluating_2020, lisle_sensemaking_2021, tahmid_evaluating_2022}, register three-dimensional content with touchscreens \cite{biener_breaking_2020, le_vxslate_2021}, and perform data visualization over existing physical elements \cite{butscher_clusters_2018, reipschlager_personal_2020, mahmood_building_2018}. VR has further been combined with a touch-screen and a spatially tracked pen in the context of authoring presentations \cite{biener2022povrpoint} and spreadsheet applications \cite{gesslein_pen-based_2020}, and to enhance capabilities of devices such as keyboards \cite{schneider_reconviguration_2019} or address privacy concerns in a mobile context \cite{grubert_office_2018, schneider_reconviguration_2019}. Further, it can reduce distractions in open office environments \cite{ruvimova_transport_2020, lee2019partitioning}, and reduce stress by experiencing a virtual nature environment  \cite{anderson2017relaxation, thoondee2017usingvirtual}.

While XR offers notable benefits with its large display space, usability challenges remain \cite{mcgill_expanding_2020, ens_personal_2014, pavanatto_monitors_2021, pavanatto_multiple_2024, ng_passenger_2021, medeiros_shielding_2022}. Low-resolution HWDs can force designers to enlarge virtual content, leading to either less screen space or more head movement \cite{pavanatto_monitors_2021}, which can result in neck pain \cite{grubert_office_2018} and reduction of task performance \cite{pavanatto_monitors_2021, pavanatto_virtual_2023}. Amplifying head rotation through virtual gains has been proposed to mitigate this issue \cite{mcgill_expanding_2020}. Context switching between physical and virtual content and focal distance changes can further decrease task performance and increase visual fatigue \cite{gabbard_effects_2019}, and induce more interaction errors \cite{eiberger_effects_2019}. In a week-long study, Biener et al. \cite{biener_quantifying_2022} observed that participants can overcome initial negative first impressions and discomfort.

While XR can shield users from physical distractions, it can also limit access to relevant real-world elements \cite{alaee2018user, desai2017window, bai2021bringing, tao2022integrating}. O'Hagen et al.~\cite{o2020reality} found that users feel uncomfortable not knowing the position of bystanders. Li et al.~\cite{li2022designing} found that HWDs led users to lower awareness of their surroundings, increased perceived workload, and lower text entry performance compared to using a tablet. Literature suggests integrating such elements into the virtual environment \cite{mcgill2015dose, hartmann2019realitycheck, wang2022realitylens}. By examining the use of VR and AR devices in public spaces, we aim to understand better how participants interact with these devices and the factors that impact their usage.

\section{Study 1: Exploring XR Work in Real-World Settings}
We conducted an exploratory user study to understand the user experience of extending a laptop screen with virtual displays when users perform their own work across real-world settings. We focused on answering: \textit{How do users take advantage of virtual displays when only a single laptop display is available?}, and \textit{How do users perceive the experience of working on virtual displays in real-world settings?} Our objective was to characterize the usage of virtual displays in such scenarios while investigating dimensions such as social, environmental, and user acceptance aspects.

\subsection{Experimental Design}
We designed this experiment such that participants completed their own productivity tasks while using a laptop screen extended with virtual displays. We observed participants in four real-world settings across the Virginia Tech campus to gather data from a wide range of contexts and to understand real-world factors. We recruited twenty participants from the student population (five for each location) who were at least 18 years old, had normal vision (corrected or uncorrected), were proficient with the English language, used a computer daily for work, and had a productivity task that they could perform during this study.

Our measures included a custom questionnaire with subjective rating scales measuring participants' perceptions of speed, accuracy, confidence, and comfort. We also obtained qualitative data through a semi-structured interview, with questions about how they perceived their productivity, what strategies they chose to use, how they felt using an AR display in public, how focused or distracted they were, and how they compared our system with the one they currently use.

\begin{figure}[tb]
 \centering
  \includegraphics[width=1.0\linewidth]{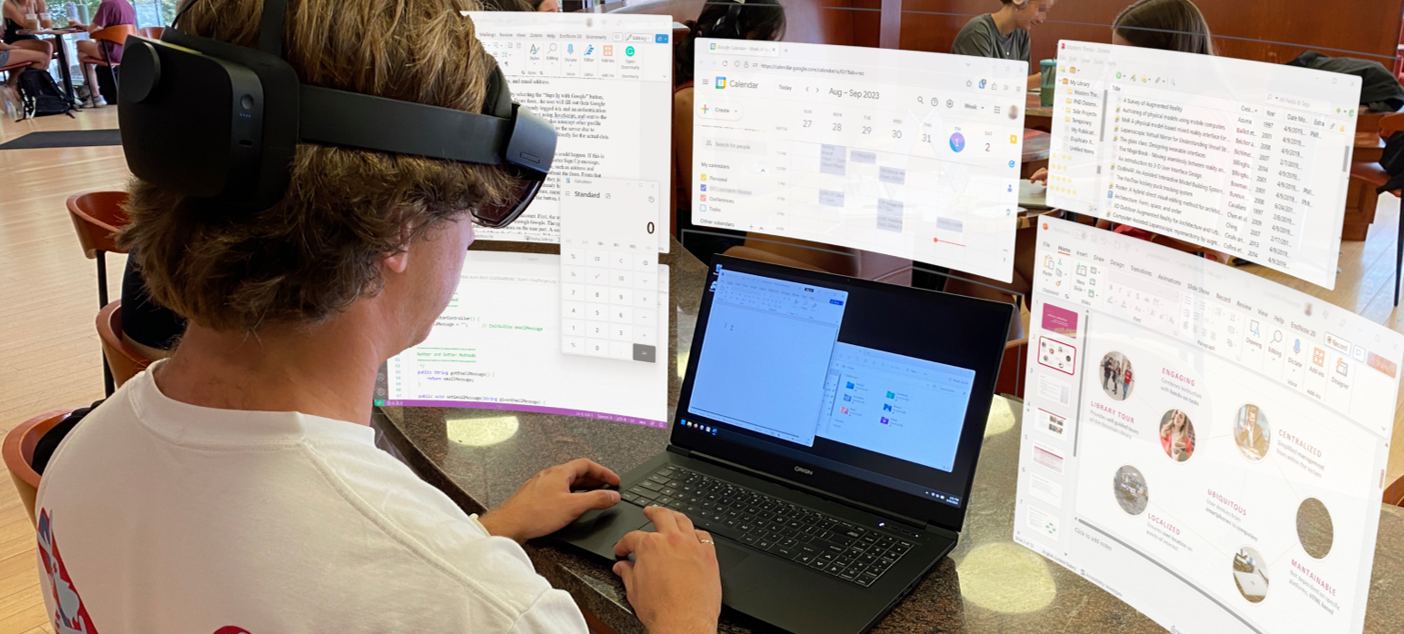}
  \caption{User working at a cafe in study 1: hybrid canvas display expands a laptop monitor with virtual space around it.}
  \label{fig:study1}
  \vspace{-0.4cm}
\end{figure}

\subsection{System and Apparatus}
We developed a \textit{hybrid canvas display} to support this work---it combines physical and virtual displays to create a large, seamless space wrapped around the left, top, and right sides of the laptop monitor \cite{pavanatto_multiple_2024}. It is curved in a semi-cylindrical shape and has a transparent background, as seen in \autoref{fig:study1}. We opted for a transparent background to allow a larger screen size without occluding the real environment---the user has the choice of where to place content so as to maintain visibility of important elements of the real world.

The laptop screen had a 2560 by 1440 pixels resolution, with 21 inches of diagonal size, and was placed about 70cm from the user. The virtual display consisted of three virtual monitors with pixel-perfect alignment and no borders between them---looking like a single display. The two side monitors had a resolution of 1920x2160, while the top monitor had a resolution of 1920x1080. All of them were scaled to 150\% to achieve higher readability. They were placed 1m away from the user---the optical sweet spot is at 2m\footnote{https://learn.microsoft.com/en-us/windows/mixed-reality/design/comfort}, but we compromised with a slightly closer placement to reduce the distance between the physical and virtual displays.

We designed our prototype using the Unity Engine 2021.3.10f1, the Mixed Reality Toolkit (MRTK) 2.8.0, and the Windows Graphics Capture API, using a back-end Visual Studio 2022 application to create and capture virtual displays. The Unity scene had curved surfaces floating in a cylinder around the user. Each surface rendered an external texture of the monitor capture obtained from the output buffer of the graphics card directly, achieving real-time display. Virtual monitors were rendered with a black background, which becomes transparent in the HoloLens. We used the Holographic Remoting Player for mirroring Windows 11 monitors on HoloLens. This application streams content from a computer to a HoloLens in real time through a tethered connection. The HoloLens sends the information obtained from its sensors (e.g., head tracking) to the PC, which makes all necessary computations and displays the result on HoloLens. For input, we used a standard laptop keyboard and trackpad across all settings; the cursor could be moved naturally across all displays without gaps between them.

We ran the experiment on an Origin Laptop (Intel i9-12900H, Geforce RTX 3080 Ti, 32GB RAM, 1TB SSD, Windows 11). We used a Microsoft HoloLens 2 HWD, with a field of view of 43\textdegree horizontally and 29\textdegree vertically, and a resolution of 2048x1080. The device's head tracking was used to register the virtual display around the laptop monitor.

\subsection{Settings}
\begin{figure*}[tb]
 \centering
  \includegraphics[width=\textwidth]{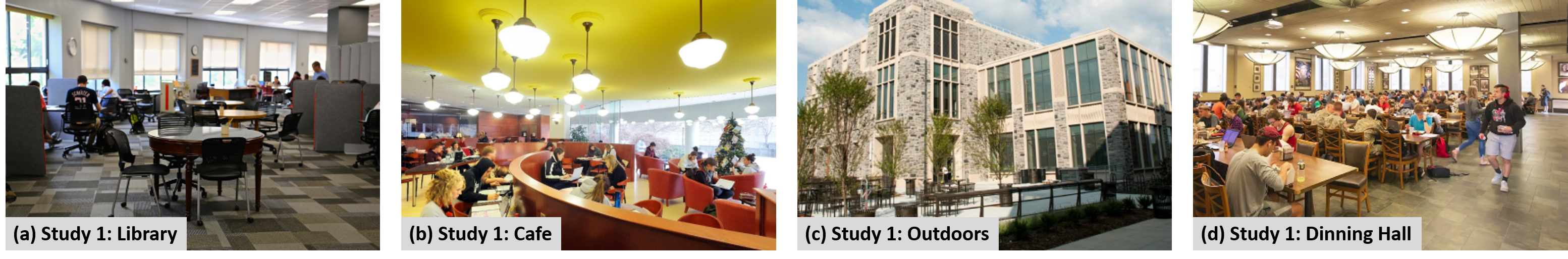}
  \caption{Study 1 was conducted across four campus settings: (a) Library, (b) Cafe, (c) Outdoors, and (d) Dining Hall.}
  \label{fig:settings_pictures}
  \vspace{-0.4cm}
\end{figure*}

We conducted our experiment across four settings shown in \autoref{fig:settings_pictures}: (a) \textbf{Library:} a less crowded public place designed for studying and working. The user can focus on the task with fewer distractions. (b)\textbf{Cafe:} a public coffee place, where people go to complete work and network. There is some noise in the environment. (c) \textbf{Outdoors:} a public outdoor space with uncontrolled illumination and temperature, where people will be working or eating. (d) \textbf{Dining Hall:} a very busy and noisy public place, where people go mostly to eat but often conduct personal work with little privacy.

These settings were chosen due to their distinct characteristics across the following dimensions: public exposure, ambient noise, brightness level, temperature level, and busyness. Two of the authors independently visited various settings across campus to characterize them. Our objective with this range of settings was to diversify user experience factors rather than to make direct comparisons between each setting.

\subsection{Task}
Upon recruitment, the investigator asked each participant about the task they would like to complete during the experiment. The task had to be productivity work that used at least three windows (to ensure that it would fit the type of multi-window work usually performed on multi-monitor setups) and used only software installed on the laptop (browsers, office suites, and coding IDEs). Browsing social media, watching videos, or doing an unclear task was not acceptable. Tasks included class assignments or side projects. In all cases, the tasks were real-world work that the participant had to complete regardless of study participation. Participants accessed their own files from the cloud or downloaded them to the computer at the beginning of the task.

\subsection{Procedure}
The study was approved by Virginia Tech's Institutional Review Board and took place in single-participant sessions of 70 minutes. Participants were recruited through mailing lists, were screened for inclusion criteria, and scheduled a session time and setting. Each participant signed the consent form and answered a background questionnaire at our laboratory. They received general instructions about the study and completed the standard HoloLens eye-calibration procedure. The participant and investigator then walked to one of the four locations (Fig. \ref{fig:settings_pictures}) where the study would take place (approximately a five-minute walk for all locations).

At the location, the investigator calibrated the virtual display to be used with the laptop. The participant put on the HWD and worked for 30 minutes on their personal task (Fig. \ref{fig:teaser} a). In case of technical issues, the experimenter was allowed to help, as we wanted the participant to use as much of the time as possible for meaningful interactions with the system. Otherwise, users were encouraged to freely explore and use the virtual displays on their own, in whichever way they felt comfortable and helpful to their workflow.

Once the work session was over, the participant answered a questionnaire with 21 rating statements on a 7-point Likert scale, focusing on subcategories such as perceived accuracy, usability, confidence, and social and environmental influence. The study ended with a semi-structured, ten-minute interview to further dive into users' usage patterns and considerations.


\subsection{Participants}
Twenty participants (aged 19 to 28, 8 female) from the campus population took part in the experiment in individual sessions of around 70 minutes. We limited our population to our university community due to the chosen settings being on campus.  One participant was a graduate student, and nineteen were undergraduates. All participants used a computer for at least 4 hours daily for work, with ten participants using it for more than 8 hours. All participants reported at least intermediate experience with the Windows operating system. Fourteen participants had little to no experience with AR.

\subsection{Methodology}
We collected our results from two main sources: background and rating questionnaires, recorded through \textit{Question Pro} \footnote{https://www.questionpro.com/}; and interview audio recordings. We performed the descriptive statistical analysis on the \textit{JMP Pro 16} software\footnote{https://www.jmp.com/}, and transcribed the interviews automatically through \textit{Office Online}\footnote{https://www.office.com/}, with manual verification and fixes completed by one of the authors. Using the transcribed interview data, one coder used a top-down approach to mark common topics in each interview using \textit{Taguette} \cite{rampin_taguette_2021}. From this initial labeling, we organized our findings into the most recurring themes. Participants were labeled as follows: initial ``L'' means the participant completed their work in the library; ``C'' means Cafe; ``O'' means Outdoors; and ``D'' means Dining Hall.

\subsection{Questionnaire Results}
\begin{figure*}[tb]
 \centering
 \includegraphics[width=\linewidth]{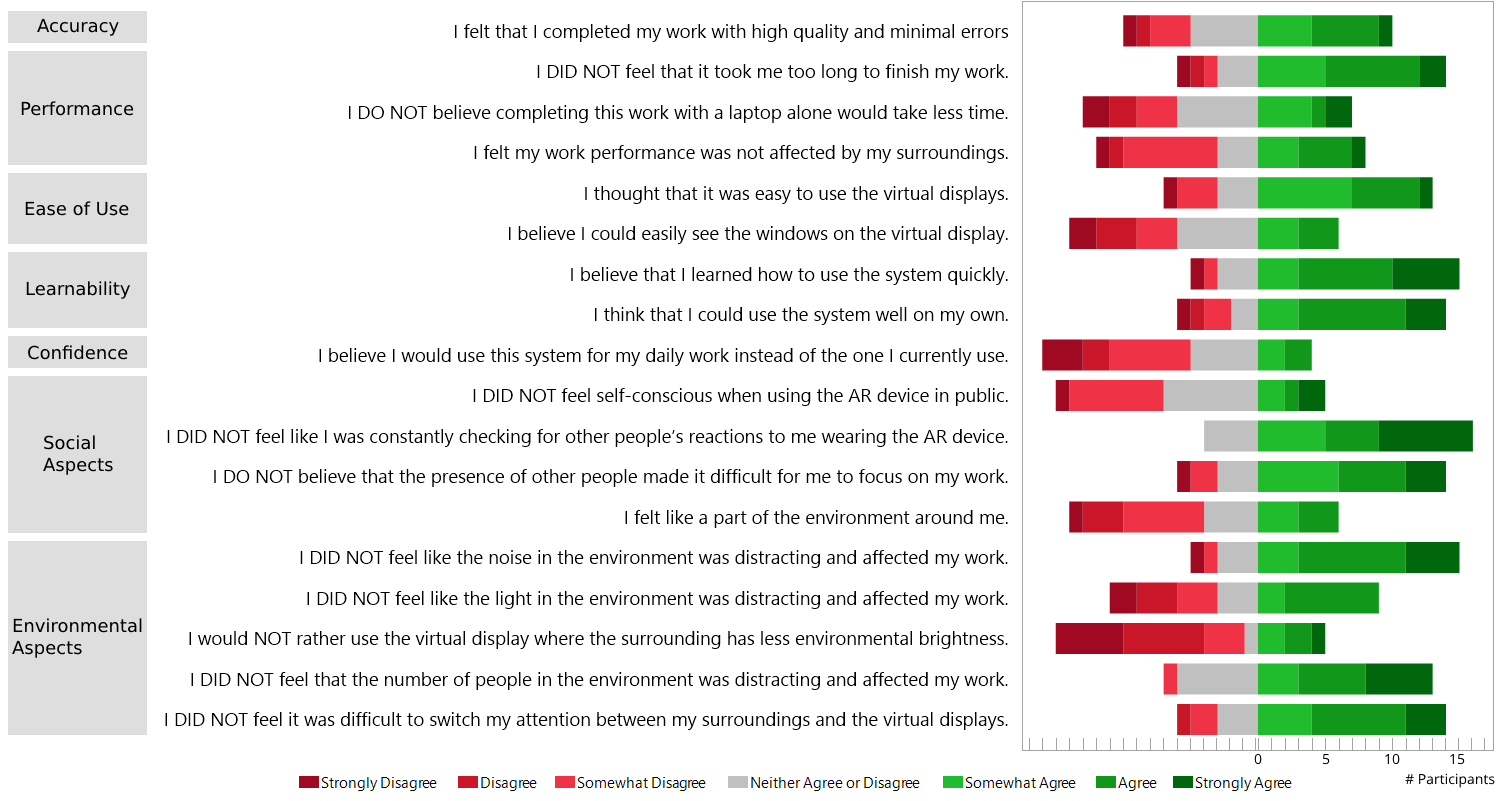}
 \caption{Questionnaire results of study 1. Higher agreement means positive results. Some of the statements were worded such that higher agreement would be negative, but we reversed the scale for those statements and added DID/DO NOT (capitalized) to them to achieve a unified scale.}
 \label{fig:questionnaireResults}
  \vspace{-0.5cm}
\end{figure*}

As can be seen in \autoref{fig:questionnaireResults}, participants overall judged the system positively on accuracy, performance, ease of use, and learnability. Regarding social aspects, they felt self-conscious but not disturbed by other people's presence. On environmental aspects, the majority of people judged that noise and the number of people around them did not affect their work, and while they adapted to existing light conditions, they preferred environments with less brightness. Finally, they did not find it hard to switch their attention between their work and surroundings.

\subsection{Usability and Utility}
\subsubsection{Similar to multi-monitor setups}
When asked about how they felt about using the system, twelve participants brought up the similarity to physical multi-monitor setups. D1 said, ``I think it was like using multiple monitors.'' O3 said, ``seemed even better than having two physical monitors because you can add a bunch of different monitors." C1 discussed the differences from a laptop: ``Definitely better than using a laptop. I'd say having multiple windows at once prevents a lot of time from switching between windows ... it was like the ability of 6 monitors.'' Respectively, L2 compared to dual-monitor setups, ``In terms of functionality, it's as good as having a second monitor.'' O4 further described benefits over traditional multi-monitors, ``You don't have to spend on a monitor and you can do all your stuff wherever you are."

\subsubsection{Support for more windows}
Expanding the laptop with virtual displays gave participants the opportunity to work with a larger number of windows than they would be able to on a laptop alone, which was mentioned in fourteen interviews. C2 was working on a coding assignment: ``I was working with multiple windows, looking at how to read Assembly code, and then reading through the Assembly to figure out how to complete the task.'' O1 said, ``I was working on an essay, and that means a bunch of research documents and resources.'' Those included the essay, personal notes, browser, and multiple papers and guidelines. L1 mentioned: ``Having the space to put all this content up was probably the biggest advantage and what I like the most about doing it in augmented reality.'' D3 compared to their laptop, ``It's definitely easier because it's nicer if you can see the assignment page and work on a different screen ... you can split your screen in a laptop too, but it gets very small."

\subsubsection{Easier access to windows}
Twelve participants highlighted their ability to quickly glance at windows. L1 was compiling a report document from various sources: ``I could see a lot more things at once. Instead of having to keep switching between tabs ... I could just look at them.'' L2 performed a comparison between documents: ``I was reading a lecture presentation and the homework answers and was comparing it with my homework.'' C1 mentioned, ``I was able to have two spreadsheets and a Google Drive folder of images open at the same time, and I could just look around to see what I needed at any given time.'' O1 mentioned ``This is nice because it's all around the center rather than having to always look over to the side, kind of losing track of the work.'' L5 added ``Usually I would have to Alt-tab back and forth ... it was nice just to see it there."

\subsubsection{Higher portability}
The lack of physical constraints of virtual displays gave thirteen participants many ideas about how they could see this system changing their daily lives. C1 said, ``I think the portability of an AR headset and the usability of multiple displays makes it easy to transport and still gives you very good functionality.'' O1 envisioned using the system when they were not near an external monitor: ``The advantage of the glasses is that I can just drag the window out of the physical monitor ... I wouldn't feel as inclined to always need an external monitor.'' O1 also mentioned spending more time working on campus, ``I'm stuck with just a tiny laptop; typically, on coding projects, researching, you have to have several windows open.'' D1 discussed the benefit of extra physical space, ``You could put so much more on your desk now that you don't have to worry about another monitor.'' L3 said ``I don't need three physical monitors, I can just use one monitor and create three virtual ones."

\subsubsection{Hardware limitations}
As reported in the literature \cite{pavanatto_monitors_2021, mcgill_expanding_2020}, we also found usability issues surrounding the use of current HWDs. Twelve participants mentioned at least one limitation. O1 wished for ``the white screens to have less rainbow effect.'' D2 said, ``A little problem is the resolution; it's a little bit too low, especially when you're writing code.'' D2 also mentioned the small field of view, ``Normally when you're looking at a real monitor, you will use your eyes to see ... in AR it only shows part of the screen.'' O3 added, ``Sometimes it was difficult to see the full window." O3 and D3 mentioned that they could see the weight becoming a problem for working for longer periods of time.

\subsection{Content Organization and Strategies}

\subsubsection{Main content placed at the center}
Fourteen participants placed their main work on the physical laptop screen. D1 said, ``My main work is in front of me, and whenever I need a reminder to look up something, I'll put it off to the side so I can glance at it and see what I need to do.'' O3 mentioned, ``Putting up text documents that I only needed to look at on the side monitors, and then on my laptop monitor were windows that I would be writing on." C1 said, ``I kept the window I was most frequently using on the laptop and supplementary windows on the outskirts in the virtual space.''  O1 agreed, ``My aim was to have the essay be the main central window, and then on the left, I had personal notes.'' L5 mentioned using the side displays for glancing, ``Part of my assignment was having to use a template ... it was nice to have the guide up there while working on an actual screen." C3 agreed, ``I kept the interactive work on the main monitor, and I put stuff that I need to reference or look at on the secondary monitors."

\subsubsection{Optimal window dimensions influence placement}
Even though the number of pixels per degree in our virtual display was much smaller than in a physical monitor, our results show that the optimal form factor for a given window can encourage users to place their main content on the virtual displays instead, which was the case for three participants. O2 opted to place their main window on the side monitor because it could be taller, displaying more information: ``I moved it over to the right where I could fill the entire rightmost display with one window, and I used the center monitor for the smaller task.'' D2, who was completing a coding assignment, said: ``The left side had a VS Code window ... especially when you look in the code, you want to look at more lines.'' L2 placed Word files they were comparing on each side of the virtual display, as ``the homework and answers were long word files, and I thought the screen was longer on the sides.''

\subsubsection{Typing tasks on virtual displays}
We expected most participants to prefer placing their typing tasks on the central monitor, given the centrality in the user view, higher resolution, and proximity to the keyboard. This was generally the case, but we did find that one participant explicitly placed their typing task on the virtual display because they wanted to experience it. L1 said, ``I wanted to see how it would feel typing on the virtual screen ... That's why I put the Google Docs that I wanted to type on the left.'' And they conclude: ``I thought typing was pretty good. The only thing is that the virtual display is still ... a little bit harder to see under certain backgrounds.'' Two other participants mentioned performing only minimal typing on the virtual displays, such as a search query.

\subsubsection{Switching windows for display quality}
One of the issues of virtual displays is dealing with limited display quality. C1 discussed this issue, as they had placed detailed images on the side displays. C1 said, ``If I really need to see details of an image I would then move that window back to my main laptop.''

\subsection{Social Factors}

\subsubsection{High awareness of the environment}
By rendering only the individual windows, and not an opaque background, in the virtual display we reduced the amount of occlusion in the environment. That led users to have a high awareness of what was happening around them, as mentioned by seven participants. L1 said, ``It was also nice to see what was going on and I could see how people were moving around me.'' O1 said, ``Right next to us, behind the window, is the restaurant self-checkout screen ... a bunch of people were coming up ... I was looking over and noticing them from time to time.'' They also mentioned that, ``When they got close enough, I decided to focus on them.'' O3 said, ``I could tell which environment I was in, but it didn't really distract me from my work more than just sitting out there with a laptop."

\subsubsection{Attention to bystanders' reactions}
Sixteen participants discussed paying attention to bystanders' reactions. C2 said, ``I could definitely see that people are looking because it is not something that you see ... but I wasn’t really thinking that much about it.'' L1 said, ``At the very beginning, there was this guy leaving and he saw me wearing the headset and I was working. He was like `ohh, what are they doing?' And I was thinking `Hmm, I don't know how I feel about that right now,' but then he left and it was fine.'' C4 said, ``You’re focusing on your work, then you're also checking who’s looking at me right now.'' O1 said, ``If someone was going to look at me, I think I understand that ... if they weren't in my immediate field of view, I don't think that would bother me. If they were close and staring at me, that'd be really awkward.'' D2 mentioned how they felt wearing the device in public, ``In the school, in the university, I think it's fine. But in a coffee shop, it would be weird.'' C1 said, ``If I saw someone using AR in public, I'd probably think, oh, that's pretty neat.'' On the questionnaire, we can further see that eight participants felt somewhat self-conscious, while another seven were neutral. At the same time, the questionnaire also revealed that none of the participants were constantly checking for bystanders' reactions to them.

\subsubsection{Easy to focus on the task}
One potential challenge of having a higher awareness of the surrounding environment would be difficulty in focusing on tasks at hand. In our study, all participants were still able to focus and complete their tasks, with fifteen commenting on how easy it was to focus on the task. D1 said ``I felt very focused ... I still can see or look around me, watch what other people do. Once I put the AR [windows] on, I actually felt relatively less self-conscious and just did my own work because my mind is more focused on the AR stuff.'' C3 said, ``I was able to focus. It was not distracting." Asked about people around them making them uncomfortable or affecting their work, L2 answered, ``Not so much. It's not 0, but it's not a lot. I was able to focus on my study.'' O5 said, ``After a while I just got absorbed into my work." D3 mentioned, ``it would be the same level of distractions if I was working on a laptop." This is reinforced by the questionnaire results: while nine participants thought the surroundings somewhat affected their work performance, thirteen reported at least partial agreement with the statement that the number of people in the environment did not distract and affect their work.

\subsubsection{Occlusion of other people in the user's view}
We observed mixed reactions to placing windows in front of other people. O1 talked about collaboration, ``If someone was in front of me working, I probably wouldn't be as inclined to put something above. That way I could still view their face ... I would not be as inclined to put one right on top of them, especially if I was working directly with them, because that would feel really weird.'' C2 didn't place windows in front of people because they knew they would be looking in that direction to view the window content: ``Just so that I wouldn't be staring at people.'' C4, D4 and D5 also followed that strategy. O2, on the other hand, said, ``The fact that it only kind of shows up on my glasses ... it didn't matter to me where things were placed.'' This sentiment was also shared by L5.

\subsection{Environmental Factors}

\subsubsection{Light background reduces readability}
A common issue with optical see-through HWDs is the reduced visibility in bright environments. This was discussed by sixteen participants, with twelve mentioning a strong impact, such as D1: ``Sometimes it felt like the screen itself was too translucent, and sometimes I felt the light was too weak.'' D5 mentioned that a yellow wall in front of them made it hard to see. C1 said, ``if it was in front of the TV at all, then it got a bit weird to see. I just kept it out of that area.'' O2 said, ``Being outside made it a little difficult. I found darker backgrounds really help.'' C5 agreed, ``I think as long as it's not in front of a light, it works fine, and you can see everything perfectly well." L1 suggested improving the system to enable a ``switch between it being transparent and not being transparent.'' This issue was further captured in our questionnaire, with fourteen participants saying they would prefer to use the display in a setting with less environmental brightness.

\subsubsection{Physical restrictions guide placement} Even though virtual displays don't need to respect the laws of physics (in that they could be placed inside another object and occlude it), four participants mentioned naturally trying to respect such restrictions. This aligns with the findings from Cheng et al. in which they found people tend to avoid occlusions when placing content in MR \cite{cheng_semanticadapt_2021}. C1 said, ``I tried to avoid putting a window below the table because visually it just looked weird to me to see a window above the table. And then just keep going through the table almost. So I try to keep all my windows above table level just so it was visually cleaner in my eyes.'' L1 used physical objects to aid their placement, ``There was a board essentially where I was looking ... so I had most of my stuff right here.''

\subsubsection{Moving backgrounds increase distraction}
One of the challenges we expected from working in AR would be users getting distracted by the environment since their view is no longer restricted by an opaque physical monitor. We observed that four participants tended to be bothered by larger people-related movements in the background, especially when they were closer to them. L1 avoided placing windows in certain regions: ``I knew that people were walking there so I didn't want to be distracted by that anyway.'' C1 suddenly stopped working at some point, and later mentioned, ``There was someone I thought I recognized for a second that distracted me for like 2 seconds.'' L2, who was in the library, was not distracted, ``At first I thought it was going to be really distracting, but it didn't eventually. So I don't think it really affected me.'' D2 follows as well, ``I think it's kind of the same experience when you just use your laptop here.''

\subsection{Discussion}
\label{discussion-study1}
Our findings demonstrate that participants had a positive acceptance of using AR virtual displays to expand a physical laptop monitor despite participants' feeling more self-conscious about wearing an HWD in public. The system was deemed similar to a multi-monitor setup, allowing participants easier access to more windows. Our results imply that virtual displays designed with current technology can be successful in accompanying existing devices in settings where additional physical monitors are unavailable.

Evidently, they also point to design challenges we should address to improve usability. While participants praised the extra space, an important aspect we uncovered was window placement being influenced by display dimensions and changing background. Coupled with user preferences, we suggest it would be more advantageous if future systems adopted an ``infinite" space on the virtual display. Since our system already curved the display over a cylinder around the user, we could instead use the entire field of regard of the user for window placement at the same fixed depth. This would provide users with further space to avoid unwanted placement areas (where people are walking by, a collaborator is standing, a light background, or a physical element such as a table), although this would also increase the amount of required physical movement, and potentially fatigue.

Another interesting aspect was the organization of content across these heterogeneous displays. Our results showed a predominance of content on the main monitor, and while there was easy access to windows in the virtual display, there was also a preference for moving content that requires high resolution to the main display. Therefore,  we must consider design strategies for allowing users to quickly switch content between the displays. Just as existing operating systems allow users to minimize content with a click, we should consider temporarily moving content to the main display (i.e., clicking or touching a button or holding down a specific key). Those would not be permanent movements but rather quick, reversible operations.

Hardware issues were noticed by participants, indicating a need for technology improvements. One specific aspect that participants reported was the need for control over window opacity. Uncontrolled settings, especially the outdoors, influence content placement due to a light background behind the display. Such issues could be reduced with video see-through HWDs, although those are subject to real-world degradation from the camera feed.

In a system able to block the outside world, we should consider interfaces that block unwanted actions that can distract workers, such as unrelated people walking far in the background. However, at the same time, systems need to be aware of the context surrounding the user, enabling easy interaction with peers and awareness of situations that may require their action. There are opportunities for research on context-aware interfaces to support such scenarios.

This study includes some limitations. First, as an exploratory study, it doesn't have a rigorous statistical analysis of some of the nuances of our findings. In addition, the sessions were limited to a single use, the HWD used in this study is still subject to various hardware limitations that should be resolved in the future, and we did not measure bystanders' perception and acceptance. Therefore, we conducted a second user study aimed at addressing those limitations, in which we fixed the setting and task across participants and instead directly compared three modalities of working (a standard laptop, an AR device, and a VR device). This allowed us to statistically analyze and compare the results of working in public with different devices while also obtaining data regarding bystanders' opinions and reactions.

\section{Study 2: User and Bystander Effects of Working in Public}
We conducted a second user study to gain insights into how users' perceptions of different measures, such as user experience, well-being, and productivity in a public space, are influenced by different interfaces, namely a standard laptop, an AR device, and a VR device. In addition, we observed reactions from bystanders and captured their opinions on using XR in public.


\subsection{Experimental Design}
The study was conducted within subjects with one independent variable \textsc{interface} having three levels, \textsc{laptop}, \textsc{ar}, and \textsc{vr}. \textsc{laptop} was chosen as the baseline, as it is common to use a laptop for working in public spaces. As discussed in section \ref{discussion-study1}, VR could have benefits over AR, such as blocking out the potentially distracting physical world, while AR provides the benefit of being aware of relevant actions in the environment. Therefore, we decided to include both types of devices in this study.  For \textsc{ar} and \textsc{vr}, we chose off-the-shelf headsets with software that can connect to a laptop as an input device while displaying multiple virtual screens---giving us the opportunity to experiment with newer and more diverse devices than in our first study. Specifically, we used the Meta Quest Pro in combination with the app “Immersed” \footnote{https://www.immersed.com/} for \textsc{vr} and the Lenovo Think Reality glasses A3 with its built-in display manager. Both devices were designed and promoted as a tool for knowledge work.

As dependent variables, we captured the perceived task load measured using the NASA TLX questionnaire \cite{hancock_development_1988}, the usability of using each device through the system usability scale \cite{brooke1996sus}, and a measure of the flow experience \cite{rheinberg2003erfassung}. Further, we decided to get an indication of participants' well-being through the simulator sickness questionnaire \cite{kennedy_simulator_1993} and by capturing visual fatigue using six questions---as done in \cite{benedetto2013readers}.
We also utilized some questions from prior studies \cite{ahlstrom2014you, koelle_dont_2015}. We pivoted to using standardized questionnaires to allow for a more rigorous and complementary analysis. 

To gain some insights into participants' emotional state, 
we asked them, ``What was your overall impression / emotion during the task?'' on a scale from 1 (I hated it, terribly awkward) to 6 (I enjoyed it, it felt comfortable) \cite{ahlstrom2014you}.  In addition, we employed five opposite word pairs as presented by Koelle et al. \cite{koelle_dont_2015} such that participants indicated their feelings on an 11-point Likert scale with the extremes being labeled with these word pairs: tense - serene, threatened - safe, unsure - self-confident, observed - unobserved, skeptic - outgoing. We counted the number of finished tasks in each condition to quantify productivity.

Finally, we again conducted a semi-structured interview after each condition, asking about the participants' behavior, what they liked and disliked, in which locations and in front of what kind of people they would be comfortable using such a system, and if they would like to use it in the future. After all conditions, we asked about their overall preference and with which device they felt most productive and most comfortable. Throughout the study, we also tracked the active window. 

In each session, we asked bystanders to fill out a short online questionnaire including demographic questions, questions about their technical affinity \cite{karrer2009technikaffinitat}, and their experience with working in public and with XR devices. In addition, they were asked whether they noticed the participant, what their feelings were about this (similar to \cite{verma2015study}), and what they thought the participant was doing. 

\subsection{System and Apparatus} 
For the \textsc{laptop} condition, we used a 16-inch HP Envy laptop (Intel i7-12700H, GeForce RTX 3060, 16GB RAM, 1TB SSD, Windows 11) with an external mouse. The laptop's resolution was kept as recommended (2560x1600 at 150\%). This laptop was also used for \textsc{vr} and \textsc{ar}. As the coloring of the keys was hard to read (white on silver), we attached stickers to the keyboard with black font on a white background to make the keys more readable in \textsc{ar} and especially in \textsc{vr}.

For \textsc{ar}, we combined the laptop with Lenovo Think Reality A3 smart glasses, connected to the laptop with a USB-C cable. Lenovo's “Virtual Display Manager 3.0.30” \footnote{https://smartsupport.lenovo.com/us/en/products/smart/arvr/thinkreality-a3/downloads/ds549422} allowed the participant to add and arrange multiple displays. We set the resolution to 1920x1080 at 125\%, which was found to be most comfortable through internal testing.

For \textsc{vr}, a Meta Quest Pro was used in combination with the app “Immersed” \footnote{https://www.immersed.com/}, which enabled a wireless connection to the laptop and then streamed the laptop content to the virtual screens. Immersed allowed participants to position the screens, choose one of nine available virtual environments, and enable pass-through windows to see the keyboard and mouse, or parts of the environment. To increase the comfort of wearing the Quest Pro, we added an additional head-strap which distributes the weight more on the center of the head.

\begin{figure*}[t]
\centering 
\includegraphics[width=1.0\linewidth]{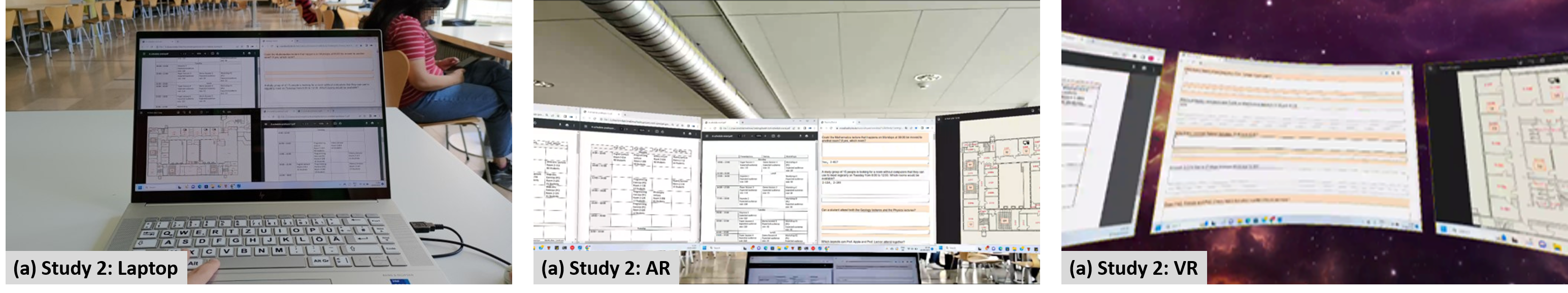}
\caption{Participant's views during the task for the laptop condition (a), AR condition (b) and VR condition (c).}
\label{fig:study2views}
\vspace{-0.4cm}
\end{figure*}

In both \textsc{vr} and \textsc{ar}, we set the number of screens to three, which could be viewed comfortably without moving the whole body and allowed distributing all the required files in a way that made them visible at the same time. Through internal testing, these settings were found to be most optimal for each condition using off-the-shelf technologies.
Both setups are depicted in Fig. \ref{fig:study2views}.



\subsection{Task}
To have a quantifiable measure of productivity, we defined a task in which progress could be measured unambiguously---as opposed to our previous study, where we focused on diversity and ecological validity. We presented a scenario of planning a schedule for university classes and a conference that would take place in the same building. Participants were provided with two PDFs, each containing the class schedules of four professors, and an additional PDF with the proposed schedule for the conference. In addition, participants had access to three floorplans images, one for each level of the building. The schedules contained information on the number of expected attendees for each class or session and whether a computer was needed. The class schedules also showed in which rooms the classes were generally held. The floor plans displayed the room numbers, capacity, and a computer capabilities icon.

Similar to the study conducted by Pavanatto et al. \cite{pavanatto_monitors_2021}, the participants got a range of questions that they needed to answer by consulting the available documents. All of these questions required the participants to view at least two different documents and compare their information to give the correct answer. An example of a task would be to figure out if a specific lecture can be moved to a different room. Therefore, the user had to check the requirements for that room in the schedule, find rooms in the floor plan that matched these requirements, and then confirm again that the room was not already booked for that time slot.


To reduce the learning effect across the three conditions, we created 3 different sets of questions and documents --- this approach was also followed by Pavanatto et al. \cite{pavanatto_monitors_2021, pavanatto_virtual_2023}.
To limit the influence of varying task difficulty, we made them comparable by using the same number of usable Rooms (30, including 10 computer rooms), the same total number of classes (90), and conference sessions (40).  The questions were similar, required similar answer lengths, and were presented to the participants in the same order.


\subsection{Procedure}
\label{sec:userstudyprocedure}
This study was conducted at Coburg University. First, the participant was asked to sign a consent form, answer demographic questions, the HEXACO questionnaire that captures certain aspects of their personality \cite{lee2004psychometric}, such as extraversion and openness to experience, and the technical affinity questionnaire (TA-EG) \cite{karrer2009technikaffinitat}. 

The study was then conducted in three sessions on separate days, one for each condition. These sessions were scheduled at the participant's convenience, yet always happened at a similar time of day to avoid lurking variables. 
We counted the number of bystanders at the beginning and end of each session and calculated the average from this, which was 19 bystanders ($sd=7$) with a minimum of 7 and a maximum of 37. The study was conducted by two experimenters.

All sessions started in a lab at the university. In the first session, the general procedure and the study task were explained to the participant, who was briefly reminded in the following sessions.
In \textsc{vr} and \textsc{ar}, they were introduced to the functionality of the devices and taught how to set up the work environment on their own and to their preferences. This included plugging in or connecting the devices, setting the IPD, and adjusting the virtual monitors (e.g., position, orientation, curvature, etc.). In \textsc{vr}, the participant was taught how to choose a virtual environment and how to add a pass-through window for seeing the keyboard and additional passthrough windows, if desired. Finally, the participant completed training to get familiar with the task and the device. 

In the cafeteria, the participant was assigned a table by the experimenter, preferably always on the same side, with the participants facing toward the bystanders, as displayed in Fig. \ref{fig:teaser} (b)-(d).
The participant sat down and set up the work environment; the experimenter sat down, positioned to get a good view of the whole room. The participant and experimenter could chat on Discord to clarify any issues. After everything was set up, the participants started with a short 5-minute task, after which they had the opportunity to adjust the setup. The main task then lasted 30 minutes, in which participants were asked to answer as many questions as possible and as accurately and completely as possible. Afterward, participants were asked to answer the questionnaires while still wearing the HWDs in \textsc{vr} and \textsc{ar}.

Throughout the study, the experimenter took notes on the events in the room and observed both the participant and other people.
After the participant finished the questionnaires, the experimenter asked several people in the room to answer a short online questionnaire accessible by scanning a QR code.

The experimenter selected bystanders to be representative of the whole room, choosing people close to the participant, people who had a good view of the participant, and people who sat further away from the participant. On average, four bystanders were asked to fill out the questionnaire in each condition. Approximately 43\% were sitting close to the participant (2 tables to left and to right, on the same or opposite side). As an incentive for filling out the questionnaire, they were offered candies. Thereafter, the participant and experimenter returned to the lab to conduct the interview. The overall process lasted, on average, two hours per session. Participants could either count this as their work time or get a gift card.

Due to technical difficulties, 3 participants had to restart the main task after 3 minutes (P24 in AR), 10 minutes (P22 in Laptop), and 15 minutes (P14 in Laptop). However, we did not exclude them as the learning effect would only be limited due to the complexity of the answers. One participant, who closed the questionnaire window by mistake less than one minute before the end, was also included.

\subsection{Participants}
Eighteen participants completed the study (5 female). Their mean age was 25.67 years ($sd=5.2$). All were students or employees at the university. All but one have studied or worked in a public space before, generally using a laptop. 
Seven had no prior experience with \textsc{ar}, 11 had at least slight experience. No one ever used \textsc{ar} in public before.
With \textsc{vr}, 10 people had slight experience, and 8 had at least moderate experience. Only one person had used \textsc{vr} in public before, which was in the university library.
They rated the frequency of using a multi-monitor setup on a scale from 1-never to 5-always. Three participants each rated their use a 1, 2 and 3, six participants rated it 4, and three participants rated it 5. Therefore, all frequencies were represented. Their average TA-EG score was 3.16 ($sd=0.54$). With a minimum of 1.85 and a maximum of 4.1 on a scale from 1 to 5.
Three additional participants were excluded due to technical problems, scheduling conflicts and pain in the eyes in the \textsc{vr} condition.

\subsection{Bystanders}
In total, we got 231 answers from bystanders. 
These answers were given by 209 different people. 
Twelve people answered questions multiple times, as they happened to be present in multiple sessions. 
We only considered the first answer someone gave for each condition. This results in 69 answers for \textsc{laptop}, 82 answers for \textsc{ar}, and 76 answers for \textsc{vr}. Of these 209 people, 104 were female (50\%), 103 male (49\%), and 2 others (1\%). Their mean age was 23.99 ($sd=4.29$) and ranged from 18 to 45, with 22 people being at least 30 years old.
When asked about their occupation, 198 said they were students, 13 named a profession in which they were working, and 2 people said they were both studying and working. Large groups of students were studying social work (60), business administration (37), computer science (30), and bioanalysis (24).

Bystanders rated their experience with each device on a scale from 1-none to 5-very much. The mean score for \textsc{laptop} was 4.1 ($sd=4.17$), for \textsc{ar} 1.34 ($sd=1.49$) and for \textsc{vr} 1.8 ($sd=2.03$).
They also rated their technical enthusiasm on a scale from 1 (not at all) to 5 (very much). The average was 2.8 ($sd=1.08$). Their technical competency was, on average, 3.08 ($sd=0.8$), their positive attitude towards technology 3.44 ($sd=0.89$), and their negative attitude 2.75 ($sd=0.89$).

All bystanders said they have used a \textsc{laptop} in public before. Only 13\% said this about \textsc{vr} devices and only 10\% have used \textsc{ar} devices in public. Locations in which bystanders have used \textsc{ar} or \textsc{vr} devices in public include universities or schools, museums, fairs, or game centers.
In contrast, 78\% stated that they saw another person using a \textsc{laptop} in public before, 33\% said they saw someone using a \textsc{ar} device in public before, and 41\% saw someone using a \textsc{vr} device in public before.
However, there is reasonable doubt that some people did not understand the question correctly, as all people answered they had used a \textsc{laptop} themselves in public, but apparently only 78\% of them saw other people using one. When asked about which devices they used to work in public, the most common answers were laptops (mentioned by 86\%), tablets (34\%), and smartphones (24\%). No one mentioned VR or AR devices.

\subsection{Methodology}
We used a repeated measures analysis of variance (RM-ANOVA) to analyze the data obtained through the study. The results can be found in Table \ref{tab:ANOVA}. For post-hoc comparisons, we applied Bonferroni adjustments at an initial significance level of $\alpha=0.05$. To ensure the robustness of the ANOVA, we used Greenhouse-Geisser correction in cases where the sphericity assumption was not met \cite{blanca2023non}.
For subjective data, we used aligned rank transform (ART) \cite{wobbrock_aligned_2011} and the ART-C method \cite{elkin_aligned_2021} for multiple comparisons, also using Bonferroni adjustments. 

We used two one-sided tests (TOST) to test the equivalence between conditions. The equivalence lower and upper bounds were calculated using Cohen's $d_z = \frac{t}{\sqrt{n}}$ with the critical $t$ value for $n=18,\alpha=0.05$ for two-tails (c.f., \cite{lakens2018equivalence}). The value of $d_z$ for our analysis was $\pm0.495$, with $-0.495$ as the lower bound and $+0.495$ as the upper bound. 
In the following we use $m$ to abbreviate the arithmetic mean and $sd$ for standard deviation.
We used axial coding to structure statements from free-text-fields and interviews.

\subsection{Results} 
\begin{table}[t]
    \centering 
    \caption{RM-ANOVA results for study 2. }
    \small
    \setlength{\tabcolsep}{5pt}

    \begin{tabular}{|c|c|c|c|c|c|}
        \hline
          & $f_{1}$ & d$f_{2}$ & F & p & $\eta^2_p$     \\ 
        \hline
        threatened vs. safe & $2$ & $34$ & $16.824$ & $<0.001$ & $0.497$   \\
        \hline
        unsure vs. self-confident & $2$ & $34$ & $4.063$ & $0.026$ & $0.193$   \\
        \hline
        observed vs. unobserved & $2$ & $34$ & $6.214$ & $0.005$ & $0.268$   \\
        \hline
        System Usability & $2$ & $34$ & $5.117$ & $0.011$ & $0.231$   \\
        \hline
        TLX Physical Demand & $2$ & $34$ & $4.686$ & $0.016$ & $0.216$   \\
        \hline
        TLX Effort & $2$ & $34$ & $5.788$ & $0.007$ & $0.254$   \\
        \hline
        Visual Fatigue & $2$ & $34$ & $15.847$ & $<0.001$ & $0.482$   \\
        \hline
        Total Simulator Sickness & $2$ & $34$  & $9.609$ & $<0.001$ & $0.361$   \\
        
        \hline
        Number of Window Switching & $1.12$ & $13.5$& $16.0$ & $0.001$ & $0.572$ \\
        \hline
        
    \end{tabular}
    
    \label{tab:ANOVA}
\vspace{-0.6cm}
\end{table}

\begin{figure*}[t]
	\centering 
	\includegraphics[width=0.9\linewidth]{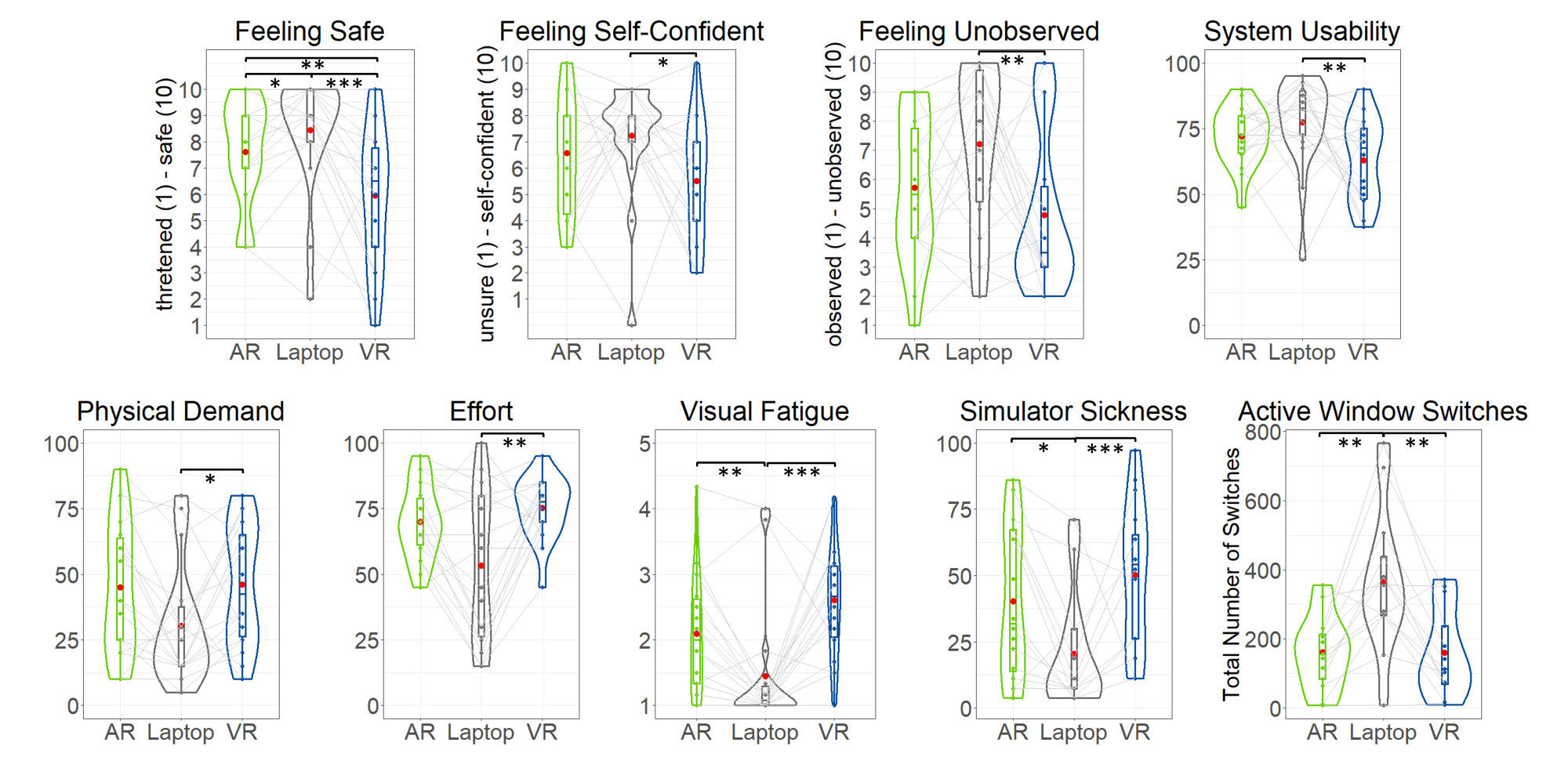}
	\caption{Violin plots for measures of study 2 where significant differences were found. Red dots indicate the arithmetic mean and stars indicate the level of significance in post-hoc tests: $*<0.05$, $**<0.01$, $***<0.001$. }
	\label{fig:SignificantDifferences}
 \vspace{-0.5cm}
\end{figure*}

\subsubsection{Being able to see the environment might increase the feeling of safety}
We found a significant main effect of \textsc{interface} on participants' rating of feeling more threatened or safe. Post-hoc tests indicate that they felt significantly more safe in \textsc{laptop} ($m=8.44,~sd=2.2$) as compared to both \textsc{AR} ($m=7.61,~sd=2.03$) ($p=0.04$) and \textsc{VR} ($m=5.94,~sd=2.67$) ($p=<0.001$) (see Fig. \ref{fig:SignificantDifferences}). In addition, participants felt significantly safer in \textsc{AR} than in \textsc{VR} ($p=0.009$), which P12 also mentioned in the interview.
Five participants mentioned they liked that they could see the real environment in \textsc{ar} 
which made them feel more secure (P5). In contrast, three participants 
didn't like not being aware of the physical environment in \textsc{vr}.
These combined findings might indicate that participants' feeling of safety is related to how well they can perceive their physical environment.
We did not observe any behavior that could have been dangerous for the participants.

\subsubsection{XR can influence self-confidence}
We also found a significant main effect of \textsc{interface} on participants' rating of feeling more unsure or self-confident. Post-hoc tests indicate that they felt significantly more self-confident in \textsc{laptop} ($m=7.22,~sd=2.18$) as compared to \textsc{vr} ($m=5.5,~sd=2.38$) ($p=0.022$) (see Fig. \ref{fig:SignificantDifferences}).
In the \textsc{vr} condition, P18 reported to feel "like a show-off".
Even though we did not find significant differences between \textsc{ar} and the other conditions, some statements from participants indicate an impact on their self-confidence.
P16 did not like to stare at a screen placed in the direction of other people, which is in line with statements from study 1 and prior work \cite{ng_passenger_2021, medeiros_shielding_2022}.
Shortly before the \textsc{ar} condition, which was her first, P7 mentioned she was nervous, and one bystander mentioned that the participant was brave for using \textsc{ar} in public.
These findings indicate that XR can have an influence on users self-confidence, especially in \textsc{vr}.

\subsubsection{Using XR-HMDs makes users stand out}
We also found a significant main effect of \textsc{interface} on participants' rating of feeling more observed or unobserved. Post-hoc tests indicate that they felt significantly less observed in \textsc{laptop} ($m=7.22,~sd=2.76$) as compared to \textsc{VR} ($m=4.78,~sd=2.67$) ($p=0.004$) (see Fig. \ref{fig:SignificantDifferences}).
P23 also stated that he felt people were looking at him in \textsc{vr}.
From observing the bystanders, we know that this feeling is reasonable. 
We saw that, unsurprisingly, no bystander seemed to be interested in the participants during the \textsc{laptop} condition, as they were blending in well with other laptop users.
During the \textsc{ar} and \textsc{vr} conditions, individual bystanders seemed to be staring at the participants for multiple seconds, some also repeatedly. However, the majority of bystanders either did not, or pretended not to notice the participants, or only looked at them in a very subtle way. Some seemed to find excuses to look such as when they walked past the participant.
While in the \textsc{laptop} condition only 65\% of bystanders indicated that they noticed the participant, this was true for 93\% in \textsc{ar} and for 95\% in \textsc{vr}.

\subsubsection{Indications of XR benefits}
Even though some measures have shown to be in favor of using a \textsc{laptop}, participants value certain aspects of the XR-experiences.
This has been indicated by the participants' choice of their overall most-liked system, in which 7 participants named \textsc{laptop}, 5 \textsc{ar}, and 6 \textsc{vr} and therefore did not show a clear preference towards one.
Most participants (12) did not like the small screen space in the \textsc{laptop} condition,
three found it annoying to switch between content 
and three were distracted by the surrounding.
Therefore, as already mentioned in study 1, ten participants liked that they had more screen space in \textsc{ar} 
which five mentioned gave them a better overview of the information 
and made them feel more productive in \textsc{ar} (P17).  
Similarly, seven participants liked the increased screen space in \textsc{vr}, 
four liked that they could be immersed in their own world, 
and five appreciated that it could be customized.
Participants also acknowledged the ergonomic benefits of \textsc{ar}, such as the possibility to look straight ahead (not down at the laptop screen) (P15, P16), which also applies to \textsc{vr}.
Specifically for \textsc{vr}, participants mentioned they could concentrate better (P8, P15, P24), as they didn't see anything else, which also made them feel more productive (P8, P24). 
But also, for \textsc{ar} P7 mentioned she noticed less from outside and P22 was less distracted than with the \textsc{laptop}.
In addition, by logging the number of active window switches (the user clicks in another window), we found a significant main effect of \textsc{interface}, even though, unfortunately, we had to exclude 5 participants due to logging errors. 
Post-hoc tests showed that in \textsc{laptop} ($m=366.62,~sd=204.76$) participants switched the active window significantly more often than in both \textsc{ar} ($m=161.54,~sd=107.08$) ($p=0.005$) and \textsc{vr} ($m=159.69,~sd=126.4$) ($p=0.004$) (see Fig. \ref{fig:SignificantDifferences}).
These observations support the findings of prior work that XR can provide certain benefits to knowledge workers. 

\subsubsection{Similar level of accuracy in all conditions}
\begin{figure*}[t]
	\centering 
	\includegraphics[width=0.9\linewidth]{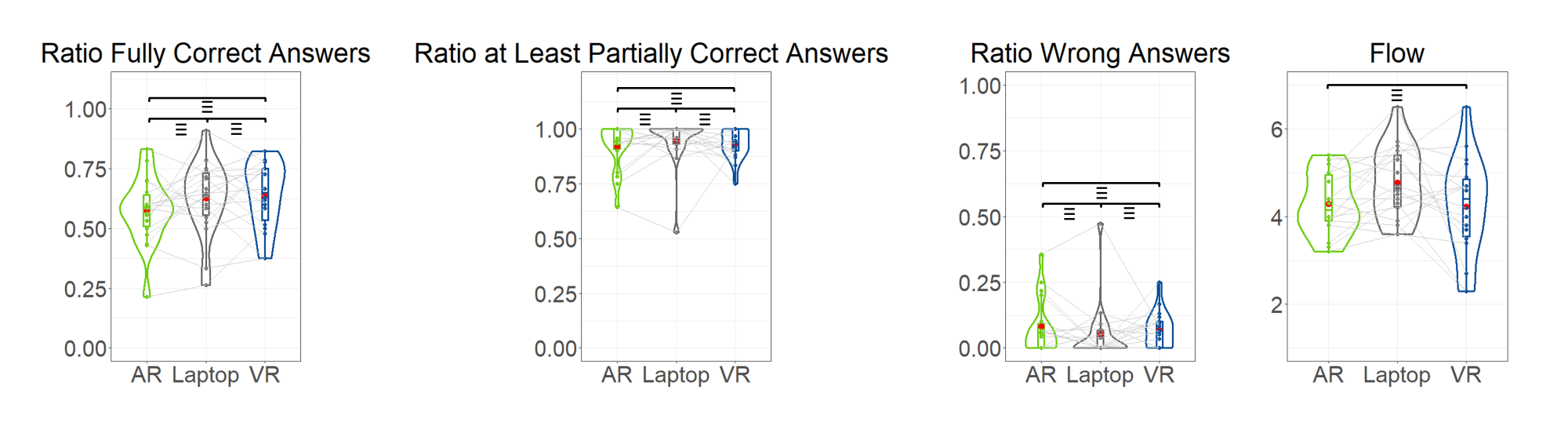}
	\caption{Violin plots for measures of study 2 where equivalences were found. Red dots indicate the arithmetic mean; equivalences are indicated by $\equiv$. }
	\label{fig:Equivalences}
 \vspace{-0.5cm}
\end{figure*}

We found an equivalence effect of \textsc{interface} on the ratios of fully correct answers between any two pairs of \textsc{laptop} ($m=0.62$, $sd=0.16$), \textsc{AR} ($m=0.58$, $sd=0.14$), and \textsc{VR} ($m=0.63$, $sd=0.14$), with both the two p-values being $p<0.001$ in all pairs. 
We also found an equivalence effect of \textsc{interface} on the ratios of at least partially correct answers between any two pairs of \textsc{laptop} ($m=0.94$, $sd=0.11$), \textsc{AR} ($m=0.92$, $sd=0.11$), and \textsc{VR} ($m=0.93$, $sd=0.07$), with both the two p-values being $p<0.001$ in all pairs. 
Similarly, the equivalence effect of \textsc{interface} was also found on the ratios of wrong answers between any two pairs of \textsc{laptop} ($m=0.05$, $sd=0.11$), \textsc{AR} ($m=0.08$, $sd=0.11$), and \textsc{VR} ($m=0.07$, $sd=0.07$), with both the two p-values being $p<0.001$ in all pairs (see Fig. \ref{fig:Equivalences}).
These findings show a similar level of accuracy between all conditions, indicating that this was not influenced by the \textsc{interface}, neither positively nor negatively.
In addition, we found an equivalence effect of \textsc{interface} on flow between \textsc{AR} ($m=2.72$, $sd=0.57$) and \textsc{VR} ($m=2.75$, $sd=0.47$), with the larger of the two p-values being $p=0.042$ (see Fig. \ref{fig:Equivalences}).
This indicates that even though participants were more isolated in \textsc{vr} this did not influence their perceived level of flow compared to \textsc{ar}.

\subsubsection{Usability of \textsc{laptop} higher than that of \textsc{vr}}
We found a significant main effect of \textsc{interface} on usability. Post-hoc tests showed that \textsc{laptop} ($m=77.36,~sd=17.11$) resulted in a significantly higher usability than \textsc{vr} ($m=62.91,~sd=15.81$) ($p=0.009$) (see Fig. \ref{fig:SignificantDifferences}).
Also, a significant main effect of \textsc{interface} was found on effort, with \textsc{laptop} ($m=53.33,~sd=29.0$) resulting in a significantly lower effort than \textsc{vr} ($m=75.28,~sd=11.82$) ($p=0.005$) (see Fig. \ref{fig:SignificantDifferences}).
In the interviews, ten participants mentioned that they liked the familiarity 
of the \textsc{laptop} condition, that they could see the keyboard (P8, P17) and type easier (P14) and that everything was compact (P7, P11). We speculate that the familiarity with the \textsc{laptop} also highly influenced the higher usability score.
In contrast, participants found \textsc{vr} slightly blurry (P6, P17), especially the pass-through (P18) so that they could not clearly see the keyboard (P18, P19). 
Usability issues reported for \textsc{ar} included losing the cursor due to the small field of view (P5, P9, P24), and the need to rotate the head to see everything (P7). Participants also mentioned the images were not always perfectly sharp (P7, P14, P23,), some noticed a flickering (P8, P14, P17) and they found the motion blur when moving the head especially annoying (P9, P16, P19, P23). As in study 1, some mentioned that a bright environment made it hard to see the virtual content (P17, P18, P22) and P21 found it distracting to see the physical screen behind the virtual screens.
However, participants liked that they could see the keyboard underneath the HMD (P8, P15, P24) in the \textsc{ar} condition, which could contribute to the slightly, yet not significantly, higher usability score, compared to \textsc{vr}.
It is noteworthy, however, that the usability scores between participants vary quite a bit, such that the group of participants who overall liked \textsc{vr} best, rated the usability of \textsc{vr} ($m=77.92,~sd=7.32$) significantly higher ($p=0.002$) than the group who preferred \textsc{laptop} or \textsc{ar} ($m=55.42,~sd=13.35$).
Similarly, P19 and P23 said they preferred \textsc{vr}, because it was easy to use while P8 found it hard to remember all settings in \textsc{vr}.
These findings point out several usability problems of both \textsc{ar} and \textsc{vr}, while only \textsc{vr} has been rated significantly worse than the baseline \textsc{laptop} condition.

\subsubsection{Current XR imposes physical discomfort}
We found a significant main effect of \textsc{interface} on physical demand, and post-hoc tests showed that \textsc{laptop} ($m=30.28,~sd=22.19$) resulted in a significantly lower physical demand than \textsc{vr} ($m=46.11,~sd=23.36$) ($p=0.025$) (see Fig. \ref{fig:SignificantDifferences}).
Also, we found a significant main effect of \textsc{interface} on the visual fatigue, with post-hoc tests showing that it was significantly lower for \textsc{laptop} ($m=1.44,~sd=0.92$) compared to both \textsc{AR} ($m=2.09,~sd=0.9$) ($p=0.003$) and \textsc{VR} ($m=2.6,~sd=0.88$) ($p=<0.001$) (see Fig. \ref{fig:SignificantDifferences}).
In addition, we found a significant main effect of \textsc{interface} on the total simulator sickness score. Again, post-hoc tests showed that it was significantly lower for \textsc{laptop} ($m=22.64,~sd=36.13$) compared to both \textsc{AR} ($m=45.71,~sd=44.06$) ($p=0.025$) and \textsc{VR} ($m=56.51,~sd=31.58$) ($p=<0.001$) (see Fig. \ref{fig:SignificantDifferences}).
This also matches the answers from participants when asked in which condition they felt most comfortable, to which 16 participants answered \textsc{laptop}, 2 \textsc{ar}, and no one \textsc{vr}. 
Participants preferred \textsc{laptop}, because they did not need to wear something (P9, P23) there was no weight on the head (P5), because no head movement was needed (P5, P9), and because it was most familiar (P8, P15) and normal (P17, P19) and did not make them feel tired (P14).
Even though the Lenovo Think Reality glasses are relatively light, participants mentioned that they pressed against their heads (P12, P23, P24), were heavy on the nose (P15), caused eye strain (P4, P14), and caused headaches (P4).
Some concluded that they could only work in \textsc{ar} for a short time (P4, P12, P23), and P4 and P12 would actually prefer it over \textsc{laptop} for a short duration.
We also observed participants supporting the HWD (P5 in \textsc{ar}), which has also been reported by Biener et al. \cite{biener2024hold}, taking breaks before starting the main task (P24) or looking troubled from the glasses (P19).
Similar answers were given with regards to \textsc{vr}, where four people had headaches, 
five had eye-problems, 
or felt slightly sick (P7) or dizzy (P15).
They also mentioned that the \textsc{vr}-HMD pressed against their head uncomfortably (P8, P14, P21, P22) and was heavy (P15, P23), or they simply didn't like to have something on their head (P5, P9).
Again, participants reported they wouldn't like to use it for long (P7, P17) and felt like they got tired faster (P12).
During \textsc{vr} conditions, we also observed that P4 was again repeatedly massaging his forehead; some adjusted the HMD (P7, P8, P11) or supported it for some time (P8, P15), similar to what was reported by Biener et al. \cite{biener2024hold}, and P21 seemed to have some pain.
These findings confirm that XR still causes physical discomfort to users, which seems to be mainly caused by the hardware specifications.

\subsubsection{Personality and experience can influence the perception of XR-use in public}
We found that the personality trait “extraversion” negatively correlates with the overall-impression in AR ($\rho(16)=-0.6,~p=0.009$). This indicates that more extroverted participants found the \textsc{ar} experience more awkward.
In addition, we found a negative correlation between AR-experience and the feeling of being unobserved in the AR condition ($\rho(16)=-0.54,~p=0.022$), and analogously, we found a negative correlation between VR-experience and the feeling of being unobserved in VR ($\rho(16)=-0.5,~p=0.037$).
This indicates that people with more \textsc{ar} or \textsc{vr} experience felt more observed in the corresponding condition.
On the other hand, the feelings of bystanders in the \textsc{vr} condition were found to positively correlate with their \textsc{vr}-experience ($r(60)=0.26,p=0.033$), indicating that with increased \textsc{vr}-experience they felt less foolish watching the participant.
These findings show that both personality and prior experience can influence how users and bystanders perceive XR use in public.

\subsubsection{Bystanders mostly feel confused or curious}
In a free-text field, bystanders were also invited to describe their feelings in more detail. Not all bystanders did this, and some expressed multiple thoughts and feelings, sometimes contradictory. 
For the \textsc{laptop} condition, 30 statements were neutral, saying the behavior of the participant was normal or they did not care.
In contrast, for the \textsc{ar} condition, only 13 statements were neutral; 15 expressed feelings like confusion or uncertainty, 15 mentioned it was cool and exciting, and another 15 found it interesting and were curious about the device. In addition, 10 were concerned with questions such as “What is this?”, “How does it work?” and “What is the person doing?”. Some bystanders expressed the wish to also try the \textsc{ar} device (4) or were wondering if they were being observed (6).
After observing the \textsc{vr} condition, only 12 statements were neutral, 23 of them expressed feelings like confusion or unfamiliarity, 13 expressed that it was exciting and amazing, and 9 mentioned it was interesting. Again, people were asking questions about the device and what the participant was doing (9) or if they could see them (3). Some again expressed the wish to try it (3).
Three bystanders explicitly mentioned they would not use \textsc{vr} in public.
One bystander said the participant looked foolish wearing the \textsc{vr} HMD, another said it was careless and two bystanders insulted the participants, calling them a loser and fool.

There were also several notable reactions. 
During \textsc{vr} conditions, 5 people looked at the participant and laughed (P4), two friends of P6 laughed at her and asked her if she could see them, two people were looking at P15 and giggling while talking to each other and a group of people looked at and talked about P16 and pointed him out to other friends.
During P5's \textsc{VR} condition, one bystander got up, saw the participant, jumped back to his friend very excited, laughed, and said, “Is this the Quest Pro?”. When passing P5 again he even waves at her. Two bystanders stared at P8 for a while and discussed if they could ask if they could try it.
In \textsc{ar}, three people sat down next to P15; as they approached, they said “Ohaa,” and one of them waved at P15.
Also, in \textsc{ar}, P15, P19, P21, and P23 were talked to for a short time.
Overall, the majority of comments were either neutral, expressing positive feelings such as curiosity or confusion with the situation.
Also, the bystanders who were more obvious with their reactions seemed rather interested than hostile, and negative comments were scarce.

Moreover, some bystanders who visited the cafeteria frequently were initially intrigued by the \textsc{ar} and \textsc{vr} headsets and observed the participants. But, over time, they did not seem curious or stare at any participants wearing those headsets, in their later visits. This could be an indication of rising acceptance of these headsets in public if people are exposed to them frequently - similar to the technology acceptance process of prior technologies \cite{edwards2003wearable, kim2009investigating}.

\subsubsection{Bystanders understanding of participant actions is less clear in XR}
We also included a text field in which bystanders could speculate about what they thought the participant was doing.
For the \textsc{laptop} condition, the most common answers were something related to studying at the university (39) or working (13).
In the \textsc{ar} condition, only 19 bystanders thought the participant was doing something study-related, and only 10 thought about working. However, 13 though the participant is playing a game. Eight were thinking about virtual worlds or content, 5 about an experiment, 4 about development and testing, and 4 thought the participant was busy with emails, videos, or news.
Four people also speculated that the participant might have eye problems. 
Regarding the \textsc{vr} condition, 25 bystanders thought the participant was doing something study-related, and 12 thought about work. This time, 20 bystanders thought the participant was playing games, 9 said they were testing something, and 4 thought about programming. Also, 6 bystanders thought the participants might be designing or constructing something. Two said they are in a virtual world, 3 said they are doing daily tasks such as videos and browsing the internet, and 3 mentioned the participant is using a virtual office, with one bystander mentioning an increased virtual screen.
These observations show that while bystanders were generally thinking that the participant was studying or working when using the \textsc{laptop}, they were often thinking about gaming when seeing \textsc{ar} and \textsc{vr}. Also, there was a wider range of guessed activities than in \textsc{laptop}.

\subsubsection{Isolation from physical surroundings}
Sometimes, bystanders were sitting down very closely to the participants. Four participants 
experienced this in \textsc{ar} but reported that they did not care. P16 was just slightly distracted when close bystander laughed.
In \textsc{vr}, P7 noticed someone sat down next to her but was wondering how close, and P12 didn't even notice someone sitting very closely. 
Through the passthrough, P5 noticed some people staring at her, but did not see the bystander who waved at her.
P19 and P21 were talked to in \textsc{ar} and reported they felt normal about it, but P21 was concerned about getting back to the task as he wanted to perform well in the study.
P24 was touched by a friend during the \textsc{vr} condition and was surprised at first, but quickly figured out who it was. This is plausible as the study took place in the university. In a completely unfamiliar environment, the participant might have reacted differently. 

Only P5 used full passthrough throughout the \textsc{vr} condition, and P9 switched to passthrough occasionally. All other participants decided to use one of the virtual environments, mostly a mountain-lodge or space-themed environment, in which they felt comfortable or which they thought looked nice.
Of the ones using a virtual environment, 14 participants did not use a passthrough window as they thought it was distracting (P4, P14, P16), was not needed (P7, P11, P17, P18, P19, P23), felt more focused and immersed (P23, P24) or felt like they had less space for the task (P12). However, P6 regretted it, as she wanted to see who was talking.
Of the three participants who used a passthrough window, P8 moved it out of sight as it interfered with the task, and P9 and P22 had one on both sides but didn't really use it during the task.
From these limited observations, it seems like participants noticed less of their environment in \textsc{vr} than \textsc{ar}, which was also reported by \cite{li2022designing}. This might also be influenced by the fact that most participants did not use passthrough windows or did not actively use them.

\subsubsection{No significant influence on productivity}
In the final interview, participants were also asked in which condition they felt most productive. Eleven participants felt most productive in \textsc{laptop}, two in \textsc{ar}, and five in \textsc{vr}.
However, we could not find a significant main effect of \textsc{interface} regarding the total given answers, the number of fully correct, partially correct, or wrong answers.
This matches findings from the first study, in which participants did not perceive their performance as better or worse than what they would have usually achieved on their own laptop.
Some participants felt most productive in \textsc{laptop} because it was more familiar (P11, P15, P22) and did not make them tired (P14).
However, P11 and P22 mentioned they could be more productive in \textsc{ar} or \textsc{vr} if they used it more often.
P17 felt most productive in \textsc{ar} because of the large screen space, and P8 felt more productive in \textsc{vr} because she could concentrate better.

\subsubsection{Preconditions for using XR in future}
Analogous to \cite{ahlstrom2014you}, we asked participants where and in front of whom they would feel comfortable using the respective devices.
Fourteen participants would feel comfortable using a \textsc{laptop} everywhere, some named limitations such as not outside (P8, P17), needing a quiet place (P19, P21) or not a location that is dynamically changing (P22). All participants said they would use the \textsc{laptop} in front of anyone.
Ten participants would use \textsc{ar} anywhere, as it is low-key (P9), they can see the environment (P14) and don't mind what others think as they don't see them (P6).
Six participants would prefer more private places, as they might feel observed (P7). 
Similar to the \textsc{laptop}, some prefer indoors (P17, P18), quiet rooms (P18, P21) and less dynamic (P22) or crowded places (P23).
Eleven participants would use \textsc{ar} in front of anybody, with exceptions such as feeling safe (P22, P23). 
Six participants prefer people who know of the device or maybe also use it 
and P12 and P24 would prefer people they know.
Only six participants would use \textsc{vr} anywhere, as they can not see the others (P4). 
Others would rather use it in only semi-public spaces such as a university (P9, P15), or in environments that provide some privacy such as a capsuled seat (P8, P16, P21). 
Some would rather use it in spaces like offices or when being alone (P6, P12, P16, P17, P18) or in stationary places (P5, P22). 
Others were concerned with being unsocial (P8) or about people's reactions (P6, P17). 
Some would prefer silent spaces (P21) or ones that are not too tight (P23).
Ten participants would use \textsc{vr} in front of anybody, under certain conditions such as feeling safe (P5). Some people would only use it in the company of people they know (P8, P12, P16, P22) or people who know about the \textsc{vr} device (P14, P18, P21).
For both \textsc{ar} and \textsc{vr}, P6 would prefer to use them in front of strangers, as friends might make fun of her.

When asked, if they could imagine using the device in their future, five people agreed for \textsc{ar}, while the other 13 could imagine it, some under preconditions such as only for a short time (P4, P14), if the field of view was larger (P5), if it was lighter (P22) and fits better with normal glasses (P18) or if it was cheap enough (P17). Some participants especially mentioned the use case of not having a fixed physical setup such as while traveling (P6, P9, P11, P15, P17, P22).
Twelve participants would like to use \textsc{vr} in the future, some of them stating some conditions such as it being more optimized (P19), only for short times (P4, P5), if it offers more advantages such as making use of the 3D display (P11), being more lightweight (P22) or being available for free (P14). Again, they mention to use it on the go (P7, P9) or to exchange reality for a better environment (P8). 
Four participants would not like to use \textsc{vr} in the future, and another two would give it another try before deciding.

These observations indicate that for both \textsc{ar} and \textsc{vr}, participants name certain requirements for places in which they would use these devices, such as semi-private rooms or less crowded or dynamic spaces. 
However, while all participants could imagine using \textsc{ar}, at least under some preconditions, 4 people could not see themselves using \textsc{vr} in the future. For both devices, participants would like to see improvements first, mainly in hardware, such as better resolution and less weight.

\subsubsection{Subtle differences between AR and VR}
As reported before, participants felt significantly safer in \textsc{ar} than \textsc{vr}.
Apart from that, participants reported that \textsc{ar} felt good (P11), was not that heavy (P21, P11), and was simple to set up (P19).
On the other hand, participants valued \textsc{vr} because there was no motion blur (P19), the screens were better (P7), and they could see more (P9).
Also, P9 found that the weight was better balanced than in \textsc{ar}, and P24 was less distracted by seeing other people. 
This could indicate that \textsc{ar} can score with a higher feeling of safety and more comfort, while \textsc{vr} provides a better visual experience with fewer distractions. 
Generally, we observed that there was a great difference between participants, with some strongly preferring \textsc{ar} and some \textsc{vr}.

\subsection{Discussion}
Our quantitative measures demonstrated several significant differences between \textsc{laptop} and XR, and our observations, in combination with participants' and bystanders' statements, gave us a more nuanced understanding of how they perceived the use of XR in public.

There were indications that participants felt safer the more they could see their physical surroundings, and participants in both studies reported liking that they could see the surroundings in AR. 
The lack of being able to see the physical world clearly also affected the usability of \textsc{vr} as some participants raised concerns about not being able to clearly see the keyboard.
On the other hand, seeing other people can also restrict the placement of virtual screens, as was also shown in prior work \cite{ng_passenger_2021, medeiros_shielding_2022}.
Most participants reported that they did not use passthrough windows in \textsc{vr} as they considered it unnecessary or even distracting. This calls for more research on how to better integrate passthrough into \textsc{vr} so that users can benefit from seeing their surroundings without distracting them or limiting the advantages of \textsc{vr}.

We also found that bystanders noticed XR users more frequently than \textsc{laptop} users. Similarly, XR users feel more observed, especially in \textsc{vr}.
However, this could change in the future if XR becomes more commonplace and more people use it.
More frequent use of XR would also increase the familiarity with the technology, which was currently found the be one of the main advantages of \textsc{laptop}. Two participants even mentioned that they think they would be more productive with XR if they used it more often.
We also observed that many bystanders were curious about the devices or even expressed the wish to try them. This could indicate that people can become open to using such systems. 
However, bystanders believed participants were studying or working in XR less often than for \textsc{laptop}. This might be because current XR technologies, especially \textsc{vr}, are more commonly known for gaming and less as a tool for knowledge work. Yet, this could already be in the process of changing, as some current XR products are advertised specifically as a tool for knowledge work.

Our findings indicate that there are still some required hardware improvements for XR to be viable for prolonged knowledge work. In particular, there is a need for a larger field of view in \textsc{ar}, which was criticized in both studies, along with the display quality and form factor, which should be improved for both \textsc{ar} and \textsc{vr}. Currently, several participants of both studies would only use the XR-devices for a short time, even though they can see benefits, such as increased screen space and fewer distractions.

We also found that personal preferences likely play an important role in whether participants preferred \textsc{ar} or \textsc{vr}, as some were very enthusiastic about the devices while others expressed a strong dislike for them.
In addition, there were indications that personality and prior XR experience can influence feelings while using XR in public or watching others doing so. Yet, we only collected a small range of data, and further research could uncover more factors that could, for example, explain preferences for certain XR devices.

Some participants, similar to study 1, reported that they might not use XR in widely open public spaces. They proposed some semi-private spaces such as offices, libraries, or places that are less dynamic and in which participants have a dedicated, maybe partially encapsulated, space, such as a seat on a train or an airplane.
Both XR devices seem to have certain advantages, such as being more aware of the surroundings in \textsc{ar} and having a slightly better screen in \textsc{vr}. Therefore, we could not find a general preference in favor of \textsc{ar} or \textsc{vr}. Choosing a device could, therefore, also depend on the situation and use case. XR devices combining both technologies could allow users to make a choice depending on their needs.
 
This study also has limitations, as the study itself could influence participants' feelings, as they might feel more confident because they are contributing to a study and not choosing to use XR devices on their own accord.
In addition, the results on productivity can heavily depend on the task, and results can therefore vary for other types of work.

\section{Design Considerations}
Overall, our results further confirm that virtual displays can be used for productivity work and that we should take advantage of their flexibility through careful interface design. 
In short, when designing virtual display systems for use in real-world settings, there are several key considerations: (1) give users unconstrained space for window placement, as users will be able to decide how their content fits better, while further avoiding unfavorable background during placements in AR; (2) physical monitors still have higher resolution and color fidelity than virtual displays, ensure that the system allows the user to take advantage of that through quick window switching or previews; (3) consider the settings where usage is most likely to occur, and consider using HWDs able to control brightness and opacity; (4) provide automatic ways of detecting background movements that should be filtered from the user; (5) be aware that XR can influence self-confidence as current HWDs still make users stand out; (6) we must consider translucency measures to bystanders, who often have no clear understanding of what headset users were doing; and (7) seeing the virtual surrounding can increase users feeling of safety.

\section{Conclusions and Future Work}
To better understand the implications of using XR in a public setting, we conducted two user studies with three different XR devices in five different public settings. We found that participants perceived XR positively, specifically highlighting the support for more windows and the high portability. In addition, we found indications that previous XR experience and personality can influence users' and bystanders' perceptions of using XR in public.  By observing and questioning the bystanders, we observed that XR still makes users stand out, and even though bystanders were generally curious about the XR devices, they did not yet have a clear understanding of their capabilities, especially for work. As conducting the studies in a university could limit the generalizability of the results, as participants and bystanders were mostly university students, future work should investigate this also in non-academic locations.


\bibliographystyle{abbrv-doi}

\bibliography{references, references-extra}




\end{document}